\documentclass[12pt]{article}

\usepackage[letterpaper,margin=3cm]{geometry}

\usepackage{authblk}

\usepackage[parfill]{parskip}
\usepackage{setspace}

\usepackage[labelfont=bf,font={small,stretch=0.95}]{caption}

\usepackage[super,sort,compress,numbers]{natbib}

\usepackage{gensymb}
\usepackage{amsmath}
\usepackage{amssymb}
\usepackage{mathtools}
\usepackage{comment}
\usepackage{hyperref}

\usepackage{graphicx}
\usepackage{multirow}

\usepackage[version=4]{mhchem}

\usepackage{mathptmx}

\usepackage{xcolor}
\usepackage{soul}
\usepackage{booktabs} 

\renewcommand{\vec}{\mathbf}

\begin{document}

\title{\Large\bf An automated framework for exploring and learning potential-energy surfaces}

\author[1]{Yuanbin Liu}
\author[1]{Joe D. Morrow}
\author[2]{Christina Ertural}
\author[1]{Natascia L. Fragapane}
\author[1]{John L. A. Gardner}
\author[2,3]{Aakash A. Naik}
\author[1]{Yuxing Zhou}
\author[2,3]{Janine George\thanks{janine.george@bam.de}}
\author[1]{Volker L. Deringer\thanks{volker.deringer@chem.ox.ac.uk}}

\affil[1]{Inorganic Chemistry Laboratory, Department of Chemistry, University of Oxford, Oxford, UK}
\affil[2]{Department Materials Chemistry, Federal Institute for Materials Research and Testing, Berlin, Germany}
\affil[3]{Institute of Condensed Matter Theory and Solid-State Optics, Friedrich Schiller University Jena, Jena, Germany}

\date{}

\maketitle

\setstretch{1.5}

{\bf
Machine learning has become ubiquitous in materials modelling and now routinely enables large-scale atomistic simulations with quantum-mechanical accuracy.
However, developing machine-learned interatomic potentials requires high-quality training data, and the manual generation and curation of such data can be a major bottleneck.
Here, we introduce an automated framework for the exploration and fitting of potential-energy surfaces, implemented in an openly available software package that we call {\tt autoplex} (`automatic potential-landscape explorer').
We discuss design choices, particularly the interoperability with existing software architectures, and the ability for the end user to easily use the computational workflows provided.
We show wide-ranging capability demonstrations: for the titanium--oxygen system, \ce{SiO2}, crystalline and liquid water, as well as phase-change memory materials.
More generally, our study illustrates how automation can speed up atomistic machine learning---with a long-term vision of making it a genuine mainstream tool in physics, chemistry, and materials science.
}

\clearpage

Machine-learned interatomic potentials (MLIPs) are now established as the method of choice for large-scale, quantum-mechanically accurate atomistic simulations \cite{Behler2017, Deringer2019, Noe2020, Unke2021, Friederich2021}, with applications ranging from high-pressure research \cite{Cheng2020, Deringer2021, Zong2021} to the discovery of molecular reaction mechanisms \cite{Westermayr2022, Zhang2024} and even to the realistic modelling of proteins \cite{wang2024ab}. MLIPs are trained on quantum-mechanical reference data---typically derived from density-functional theory (DFT)---using a variety of methods from linear fits \cite{Thompson2015, Shapeev2016, Drautz2019} and Gaussian process regression \cite{Bartok2010} to neural-network architectures\cite{Behler2007, ZhangPhysRevLett.120.143001, unke2019physnet, Batzner2022, MACE2022, Chen2022, Deng2023}. 
Traditionally, MLIPs have been largely hand-crafted models, built using configurations manually tailored for domain-specific tasks \cite{Sosso2012, Bartok2018, Erhard2022, zhou2023device}, such as the fracture of silicon \cite{Bartok2018} or the crystallisation of Ge--Sb--Te memory materials \cite{zhou2023device}. More recently, a trend has emerged towards pre-trained or `foundational' MLIPs \cite{zhang2024dpa2largeatomicmodel, Batatia2023}: these models are fitted to large datasets including many chemical elements, and can be fine-tuned for downstream tasks\cite{zhang2024dpa2largeatomicmodel, Kaur2024}. 

With sophisticated MLIP fitting frameworks available and continuously improving, we argue that the next area of innovation lies in the data used to train the models \cite{BenMahmoud2024, kulichenko2024data}. The aforementioned fine-tuning is one example of the more general challenges in this field: constructing high-quality datasets still remains a non-trivial, time- and labor-intensive aspect of MLIP model development, and (more) efficient methods for data generation are needed. Commonly, active-learning strategies are now used to iteratively optimise datasets by identifying rare events and selecting the most relevant configurations via suitable error estimates\cite{smith2018less, van2023hyperactive, kulichenko2023uncertainty}. Active learning has been widely used to explore phase transitions \cite{vandermause2020fly, vandenhaute2023machine, xie2023uncertainty} and chemical reactions\cite{young2021transferable, guan2023using, schaaf2023accurate}.
And yet, such methods often still rely on costly {\em ab initio} MD computations to expand and refine the training datasets. 

\begin{figure*}
    \centering
    \includegraphics[width=\linewidth]{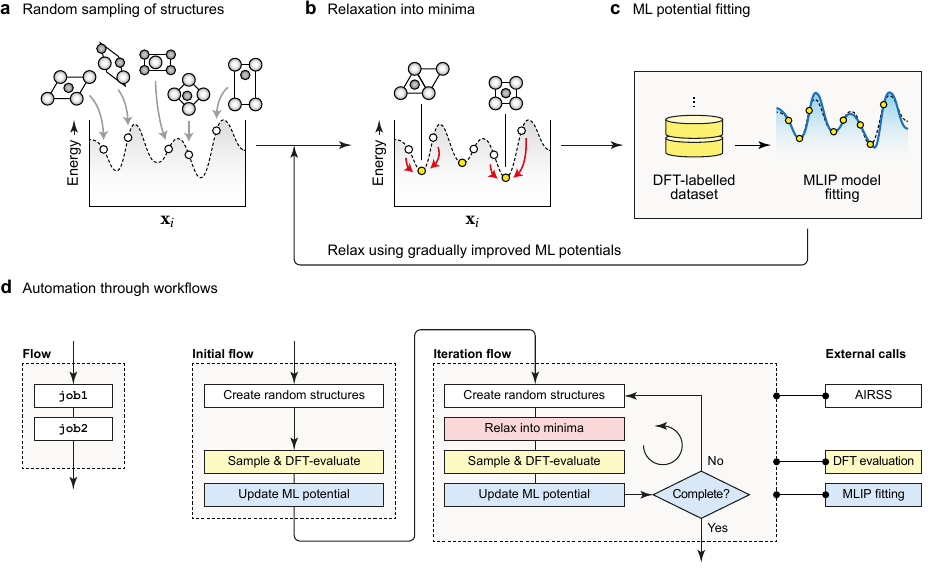}
    \caption{\textbf{Exploration and fitting of potential-energy surfaces.} 
    From left to right, the process involves: ({\bf a}) the random sampling of structures, using suitable constraints (in this simple cartoon, we assume that we are searching for crystal structures of a binary compound, containing one atom of each type under periodic boundary conditions); 
    ({\bf b}) the relaxation of those structures into local minima which correspond to relevant crystal structures; and ({\bf c}) the fitting of ML potentials. 
    Panel ({\bf d}) provides a high-level overview of the workflow structure: the left-hand side is a cartoon of a `flow' consisting of two arbitrary `jobs' to be carried out in sequence, and the schematic in the main part of the figure summarises our approach taken in {\tt autoplex}. We colour-code the different key steps in {\em white} (random structure generation, here using the {\tt buildcell} code of AIRSS \cite{Pickard2006, Pickard2011}), {\em red} (iterative exploration), {\em yellow} (DFT single-point computations), and {\em blue} (ML interatomic potential fitting).}
    \label{fig:overview}
\end{figure*}

Currently available `foundational' MLIPs are typically fitted to a dataset comprising relaxation trajectories of diverse crystalline materials sourced from the Materials Project initiative \cite{Jain2013}. The present work is concerned with the---somewhat orthogonal---question of how one can build an MLIP model from scratch: exploring local minima but also highly unfavourable regions of a given potential-energy surface, which need to be taught to a robust potential. Previous work showed that random structure searching (RSS) provides a particularly promising approach for the exploration and iterative fitting of configurational space (Fig.~\ref{fig:overview}a--c)\cite{Deringer2018, Tong2018, Deringer2018c, Podryabinkin2019, Bernstein2019b, Pickard2022, Pickard2024}. The principle of the original RSS approach, known as {\em Ab Initio} Random Structure Searching (AIRSS)\cite{Pickard2006, Pickard2011}, is illustrated by panels a--b in Fig.~\ref{fig:overview}, and the GAP-RSS approach proposed in Ref.~\citenum{Deringer2018} unifies it with MLIP fitting: using gradually improved potential models to drive the searches, without relying on any first-principles relaxations (only requiring DFT single-point evaluations) or pre-existing force fields.
We note that AIRSS has been used as part of the approach for developing the recently described `graph networks for materials exploration' (GNoME) \cite{Merchant2023} and MatterSim \cite{Yang2024} models to create structurally diverse training data. 

To date, these RSS-related approaches still depend heavily on the user's expertise and time and are by no means trivial to implement. This challenge is particularly apparent for very large training datasets, where manually executing and monitoring tens of thousands of individual tasks is practically impossible. A similar challenge was previously observed in DFT-driven materials discovery, and automation approaches have been developed in response \cite{George2021a}: numerous workflow systems can now be used to streamline first-principles materials exploration \cite{curtarolo2012aflow, kirklin2015open, pizzi2016aiida, mathew2017atomate, choudhary2020joint}. Owing to such efforts, together with high-performance computing facilities, DFT-driven high-throughput simulations have become commonplace today and have played an important role in the computational discovery of new materials\cite{Pyzer2015, gorai2017computationally, HAUTIER2019108}. However, the same level of maturity has not yet been achieved for the full development pipeline of MLIPs (exploration, sampling, fitting, refinement): although important steps have recently been made\cite{JANSSEN_pyiron_201924, DP-GEN_ZHANG2020107206, gelvzinyte2023wfl, Guo_2023_10.1063/5.0166858, Menon2024, Poul2024}, the development of fully automated workflows for MLIPs remains in high demand.

Here, we describe an automated implementation of iterative exploration and MLIP fitting through data-driven random searching, focusing on both the automation infrastructure and its implications for materials modelling applications.  
We show that, within the open-source {\tt autoplex} code we have developed, MLIP fitting can be carried out in a largely automated way on high-performance computing systems and in a high-throughput manner---and we show how the resulting potentials are robust and useful, especially given the ease with which they can be created from scratch.
We expect that this work will contribute to the mainstream uptake of ML-driven atomistic modelling in the wider community in the years ahead.

\section*{Results}

\subsection*{The {\tt autoplex} framework}

The {\tt autoplex} framework is a modular set of software that is interfaced to widely-used computational and automation infrastructure where applicable.
In particular, our code follows the core principles of (and reuses some functionalities implemented in) the {\tt atomate2}\cite{ganose_atomate2_2024} framework, which in turn underpins the Materials Project\cite{Jain2013} initiative. 
The code is available openly via GitHub, distributed under a permissive licence, and accompanied by documentation to facilitate its uptake. In the present work, we mainly use the Gaussian approximation potential (GAP) \cite{Bartok2010} framework to drive exploration and potential fitting, leveraging the data efficiency of GAP and building on previous successful applications for this purpose \cite{Deringer2018, Bernstein2019b}. We do note that {\tt autoplex} is designed to accommodate other MLIP architectures as well.

\begin{figure*}
    \centering
    \includegraphics[width=\linewidth]{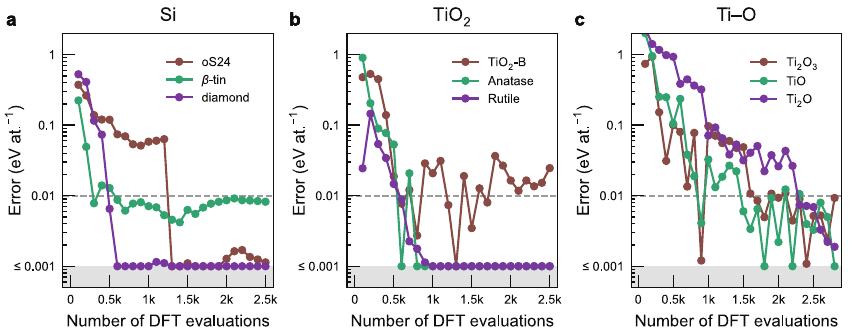}
    \caption{{\bf Automated exploration of key materials systems.} We show the results of {\tt autoplex} runs, characterised by the energy prediction error of the models, for gradually more complex scenarios: ({\bf a}) the diamond-type and $\beta$-tin-type form of elemental silicon, as well as the open-framework oS24 allotrope described in Ref.~\citenum{Kim2015}; ({\bf b}) the common rutile and anatase forms of \ce{TiO2}, as well as the less common `B' polymorph; and ({\bf c}) three phases from the full binary Ti--O system. Note that the exploration for panel (b) included only stoichiometrically precise \ce{TiO2} structures, whereas that in panel (c) used a variable composition range. The data are shown in a style similar to Ref.~\citenum{Bernstein2019b}, with energy errors limited by a set minimum value (we take 0.001 eV at.$^{-1}$ to roughly correspond to the accuracy limit achievable using DFT \cite{Bernstein2019b}). Lines between symbols are guides to the eye.}
    \label{fig:si-tio}
\end{figure*}

We demonstrate the validity of the method using key examples, moving up in difficulty from elemental silicon to the polymorphs of \ce{TiO2}, and onwards to the full binary titanium--oxygen system. Figure \ref{fig:si-tio} shows the evolution of energy prediction errors for relevant crystalline modifications with an increasing number of DFT single-point evaluations used to create GAP-RSS models. Each panel in Fig.~\ref{fig:si-tio} corresponds to a separate round of iterative, automated training using {\tt autoplex}. Each step encompasses 100 single-point DFT evaluations whose results are added to the training dataset.

Silicon (Fig.\ \ref{fig:si-tio}a)---perhaps the classic test case for any materials simulation method---has a main allotrope with the diamond-type structure, and multiple higher-pressure forms, prominently the $\beta$-tin type structure. We also include the open-framework oS24 allotrope of silicon \cite{Kim2015} as an example of a metastable phase that has been experimentally characterised and includes lower-symmetric atomic environments, as a more challenging test case for {\tt autoplex}.
The tests in Fig.\ \ref{fig:si-tio}a are similar to earlier work in Ref.\ \citenum{Bernstein2019b} and, for consistency, use the same DFT parameters as in that previous work (e.g., the same exchange--correlation functional). All three allotropes are well described to within an accuracy on the order of 0.01 eV at.$^{-1}$: the highly symmetric diamond- and $\beta$-tin-type structures with $\approx$ 500 DFT single-point evaluations, the oS24 structure within a few thousand (Fig.\ \ref{fig:si-tio}a). We take 0.01 eV at.$^{-1}$ to be a `sensible' accuracy target for random exploration, and indicate this value by dashed horizontal lines in Fig.\ \ref{fig:si-tio}. The higher numerical error for $\beta$-tin-type compared to diamond-type silicon (green vs.\ blue in Fig.\ \ref{fig:si-tio}a) is consistent with previous work on a general-purpose GAP model for the element \cite{Bartok2018} as well as an earlier GAP-RSS study \cite{Bernstein2019b}.

The binary oxide, \ce{TiO2}, is structurally highly diverse, therefore forming a suitable next target for testing a crystal-structure searching method. The compound mainly exists in two common forms---rutile and anatase---which contain distorted octahedrally coordinated [\ce{TiO6}] units and differ in the connectivity of the coordination polyhedra. We also include the bronze-type (`B-') polymorph of \ce{TiO2}, which is less abundant, but has been of interest for battery research \cite{Armstrong2005, Liang2022}.
Figure \ref{fig:si-tio}b shows that while the two main polymorphs are again correctly recovered, and the prediction error for B-\ce{TiO2} reduces to a few tens of meV at.$^{-1}$ as well, the latter polymorph appears to be distinctly more difficult to `learn' than the two simpler ones.

\begin{table}[t]
    \caption{Comparison of errors between the GAP-RSS-based \ce{TiO2} potential (Fig.~\ref{fig:si-tio}b) and the Ti–O potential (Fig.~\ref{fig:si-tio}c) across major Ti–O polymorphs. Each error estimate is based on 10 `rattled' structures, generated by applying random displacements to the atomic positions of the ground-state structures with a standard deviation of 0.01 \AA{}.}
    \centering
    \setlength{\tabcolsep}{5.0mm}{
    \begin{tabular}{llcccc}
    \hline
    \hline
    & & \multicolumn{2}{c}{RMSE (meV at.$^{-1}$)} \\
    \cline{3-4}
    & & GAP-RSS & GAP-RSS \\
    Compound & Structure type & (\ce{TiO2} only) & (Full Ti--O system)\\
    \hline
    \ce{TiO2} & Anatase & $0.1$ & $0.7$ \\
    \ce{TiO2} & Baddeleyite & $1.1$ & $28$ \\
    \ce{TiO2} & Brookite & $10$ & $8.2$ \\
    \ce{TiO2} & Columbite & $1.0$ & $0.9$ \\
    \ce{TiO2} & Rutile & $0.2$ & $1.8$ \\
    \ce{TiO2} & \ce{TiO2}-B & $24$ & $20$ \\
    \hline
    \ce{Ti3O5} & \ce{Ti3O5} & $105$ & $19$ \\
    \ce{Ti3O5} & \ce{V3O5(HT)} & $10$ & $4.1$ \\
    \ce{Ti2O3} & \ce{Al2O3} & $144$ & $9.1$ \\
    \hline
    \ce{TiO} & NaCl & ---$^{a}$ & $0.6$ \\
    \ce{Ti2O} & \ce{Ti2O} & ---$^{a}$ & $2.2$ \\
    \ce{Ti3O} & \ce{Ti3O} & ---$^{a}$ & $23$ \\
    \hline
    \hline
    \multicolumn{4}{p{12cm}}{\footnotesize $^{a}$In these cases, errors were $> 1$ eV at.$^{-1}$, and numerical values are therefore not meaningful to report.} \\
    \end{tabular}}
    \label{tab:TiO}
\end{table}

We finally study the exploration of a full binary system containing multiple phases with varied stoichiometric compositions.
In Fig.~\ref{fig:si-tio}c, we present the results of testing our approach on compounds with different stoichiometric compositions (and electronic structure), viz.\ \ce{Ti2O3}, \ce{TiO}, and \ce{Ti2O}. While we truncate the plot at 0.001 eV at.$^{-1}$, we emphasise that we already consider achieving an accuracy of $\approx 0.01$ eV at.$^{-1}$ to be sufficient in this test of random exploration. It is fair to observe that compared to simpler phases such as rutile and anatase, achieving the target accuracy in this case requires a greater number of iterations,  as the search space is more complex.

Table \ref{tab:TiO} shows results for relevant main Ti--O polymorphs, evaluated with the \ce{TiO2} potential from Fig.\ 2b, and with the Ti--O potential from Fig.\ 2c. 
This is a particularly instructive case because it allows us to probe the limits of the method: if only trained on \ce{TiO2}, a GAP-RSS model will faithfully capture the polymorphs with this specific stoichiometric composition, but produces unacceptable errors when applied to compositions that deviate largely from the stoichiometry ($>100$ meV at.$^{-1}$ for one of the \ce{Ti3O5} polymorphs, and $>1$ eV at.$^{-1}$ for rocksalt-type TiO, for example). In contrast, by training the model for the full Ti--O system (cf.\ Fig.\ \ref{fig:si-tio}c), we are able to obtain an accurate description for several different phases. This example highlights the flexibility of {\tt autoplex} in handling varying stoichiometric compositions, requiring no substantially greater effort from the user than that for a single stoichiometrically precise compound---all that is required is a change in input parameters for RSS, and probably a moderately increased amount of computational resources. In contrast, the workload for common manual (AI)MD-based approaches would increase substantially, as a single trajectory typically handles only one crystal type and stoichiometric composition. We note that likely, the overall accuracy of those models can be improved further---for the time being, we report them as obtained with standard DFT and GAP fitting settings.

\subsection*{High-level potentials at moderate computational cost}

A distinct advantage of the RSS approach for potential fitting is that it requires only single-point computations to generate the reference data\cite{Deringer2018}. We therefore posit that we are able to use {\tt autoplex} to easily build high-quality potentials beyond the `standard' GGA functionals that are commonly used. We show here an example where higher-level data are crucial.

\begin{figure*}
    \centering
    \includegraphics[width=\linewidth]{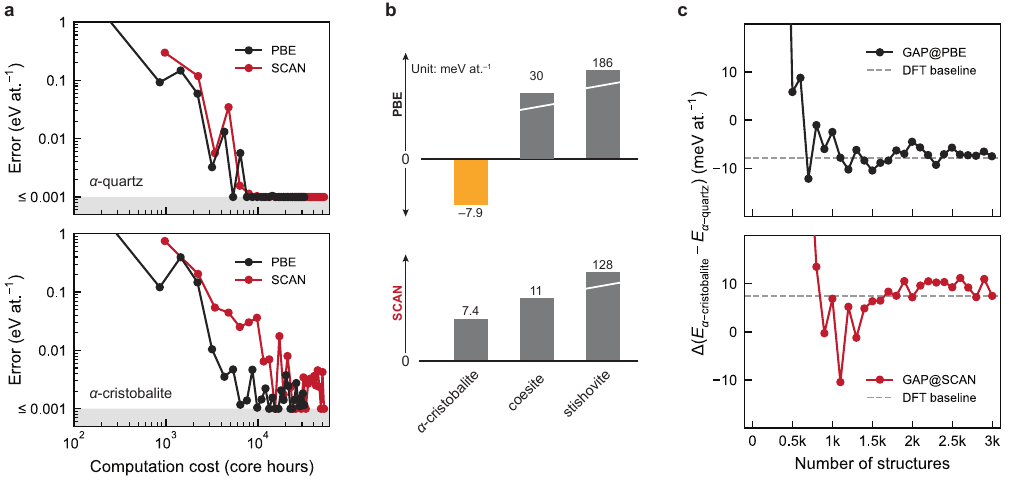}
    \caption{{\bf Accessing higher-rung DFT for \ce{SiO2}.} ({\bf a}) Energy prediction errors for $\alpha$-quartz and $\alpha$-cristobalite, plotted as a function of computational cost for {\tt autoplex} runs using the PBE and SCAN functionals, respectively. ({\bf b}) DFT-computed energies of key \ce{SiO2} polymorphs, relative to $\alpha$-quartz which is set as energy zero. Note that PBE erroneously predicts $\alpha$-cristobalite as being more stable than $\alpha$-quartz. ({\bf c}) Energy difference between $\alpha$-quartz and $\alpha$-cristobalite as a function of the number of training structures for GAP@PBE and GAP@SCAN models.}
    \label{fig:sio2}
\end{figure*}

The seemingly simple silicon dioxide, \ce{SiO2}, has long posed challenges for atomistic modelling (see Ref.~\citenum{Erhard2022} and references therein). We start by running our workflow using the economic Perdew--Burke--Ernzerhof (PBE) functional \cite{Perdew1996}. We then re-run the workflow with the same RSS parameter settings, this time using the Strongly Constrained and Appropriately Normed (SCAN) functional \cite{Sun2015}. Figure \ref{fig:sio2}a shows that for $\alpha$-quartz, achieving high prediction accuracy (1 meV at.$^{-1}$) with both PBE and SCAN functionals requires less than 10,000 CPU core hours, corresponding to nominal costs on the order of \$100. For the structurally more complex $\alpha$-cristobalite polymorph, SCAN incurs higher computational costs compared to PBE but still remains at the scale of 10,000s of core hours.

Why SCAN? The importance of using this higher-rung functional in this case becomes apparent when inspecting the {\em absolute} energy predictions of the different MLIPs. PBE fails to correctly reproduce the stability ordering of \ce{SiO2} polymorphs (see Fig. 3b), incorrectly predicting that $\alpha$-cristobalite is more stable than $\alpha$-quartz. This is an issue that is well known for some DFT methods (see, e.g., an early study in Ref.~\citenum{Demuth1999}). By contrast, SCAN does identify $\alpha$-quartz as stable---and indeed this DFT level has been used to train an MLIP for \ce{SiO2} before \cite{Erhard2022}, as well as a more comprehensive one for the binary Si--O system \cite{Erhard2024}. Further work could now include a direct benchmarking of our candidate RSS-derived MLIPs, as well as other data-generation workflows \cite{Guo_2023_10.1063/5.0166858, gelvzinyte2023wfl, Menon2024, Poul2024}, against the models of Refs.~\citenum{Erhard2022} and \citenum{Erhard2024}. 

Figure 3c illustrates the energy difference between $\alpha$-cristobalite and $\alpha$-quartz as a function of the number of training structures for GAP models trained with PBE and SCAN. The results indicate that the stability predictions from both GAP-RSS runs align well with the respective DFT results---but because of the underlying training data, the GAP@SCAN model is qualitatively correct ($\Delta E > 0$), whereas the GAP@PBE model is not ($\Delta E < 0$).

Table \ref{tab:SiO2} shows the results for additional \ce{SiO2} polymorphs: aside from $\alpha$-cristobalite, the stability order of moganite and tridymite with respect to $\alpha$-quartz is also inaccurately predicted by DFT@PBE and GAP@PBE. In contrast, DFT@SCAN and GAP@SCAN successfully capture the stability of different crystal structures with overall accuracy. However, in terms of numerical precision, there is no significant difference between the performance of the GAP@PBE and GAP@SCAN models. Even polymorphs with more complex unit cells are still predicted within our primary accuracy target of 10 meV at.$^{-1}$.

\begin{table}[t]
    \caption{Comparison of energy differences between various \ce{SiO2} polymorphs and $\alpha$-quartz using different DFT methods and GAP models at the PBE and SCAN levels. (Unit: meV at.$^{-1}$)}
    \centering
    \setlength{\tabcolsep}{5.0mm}{
    \begin{tabular}{lcccc}
    \hline
    \hline
    & \multicolumn{2}{c}{$\Delta (E - E_{\alpha\text{-quartz}})_{\text{PBE}}$} & 
    \multicolumn{2}{c}{$\Delta (E - E_{\alpha\text{-quartz}})_{\text{SCAN}}$} \\
    \cmidrule(lr){2-3} \cmidrule(lr){4-5}
    Structure type & DFT & GAP-RSS & DFT & GAP-RSS\\
    \hline
    coesite & 30 & 31 & 11 & 12\\
    stishovite & 186 & 185 & 128 & 127\\
    \hline
    \textit{$\alpha$}-cristobalite & {\bf --7.9}$^{a}$ & {\bf --7.5}$^{a}$ & 7.4 & 7.5\\
    moganite & {\bf --0.4}$^{a}$ & {\bf --3.5}$^{a}$ & 1.8 & 6.2\\
    tridymite & {\bf --8.2}$^{a}$ & {\bf --11}$^{a}$ & 8.4 & 7.6\\
    \hline
    \hline
    \multicolumn{5}{p{12cm}}{\footnotesize $^{a}$These polymorphs are erroneously predicted to be more stable than $\alpha$-quartz at the PBE level.} \\
    \end{tabular}}
    \label{tab:SiO2}
\end{table}

\subsection*{Describing water with different architectures}

While most of the potentials herein use the GAP framework, we explore the use of different MLIP fitting approaches, specifically that of graph-neural-network potentials. We do this using a `textbook' example of a molecular system.

\begin{figure*}
    \centering
    \includegraphics[width=0.75\linewidth]{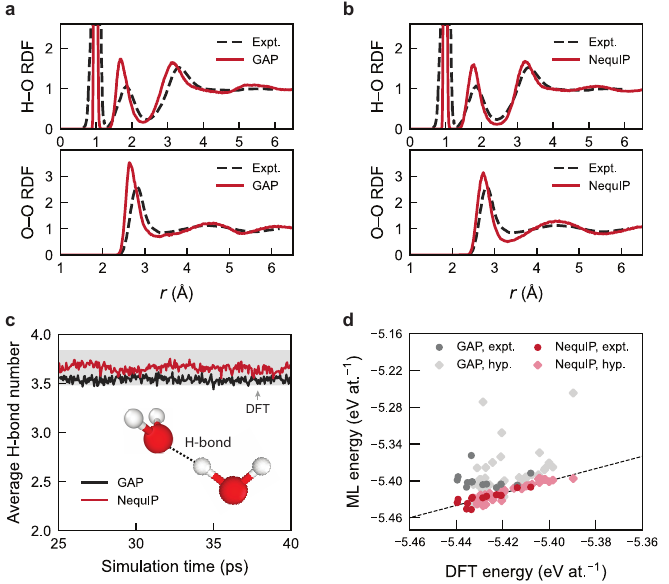}
    \caption{\textbf{Liquid water and ice polymorphs.} ({\bf a}) Radial distribution functions (RDFs) of H--O and O--O pairs comparing experimental data (H--O: Ref.~\citenum{Soper2000}; O--O: Ref.~\citenum{Skinner2014}) with GAP-RSS model predictions at 300 K.
    ({\bf b}) Same for predictions from a NequIP model, fitted to the same training set as the GAP-RSS model. ({\bf c}) Average hydrogen-bond number as a function of simulation time for GAP and NequIP models, showing close agreement with previous DFT-calculated values (data from Ref.~\citenum{Todorova2006}). The hydrogen-bond criterion is defined as the O–O distance being smaller than 3.5 ~\AA~ and the O–H–O angle being larger than 140$^\circ$. ({\bf d}) Comparison of ML-predicted energies from GAP and NequIP models fitted to the GAP-RSS dataset, against DFT energies for various ice polymorphs \cite{monserrat2020liquid}, including those experimentally observed (labeled `expt.') and those theoretically hypothesised (labeled `hyp.'). For this test, structures were taken from Ref.~\citenum{monserrat2020liquid}, and re-evaluated with DFT using the SCAN functional \cite{Sun2015}. This plot highlights the improved extrapolation capability of the NequIP model, compared to the initial GAP-RSS one, in capturing energy trends across different phases.
    }
    \label{fig:water}
\end{figure*}

Figure \ref{fig:water} characterises results for liquid water and ice polymorphs, obtained using the SCAN functional \cite{Sun2015}. We run GAP-RSS iterations as before, initially including configurations of $> 1$ eV at.$^{-1}$ above the convex hull, and then gradually exploring lower-energy structures. These searches typically use small unit cells, and therefore it is encouraging to see that the resulting potentials lead to a qualitatively correct description of the structure of liquid water in MD simulations (Fig.\ \ref{fig:water}a). We do not present the H$\cdots$H interactions here, as our MD simulations do not account for nuclear quantum effects, which significantly influence the first peak of the radial distribution function (RDF)---see, e.g., Ref.~\citenum{Markland2018} for a discussion of these effects.

With an initial GAP-RSS-based dataset available, we carried out fits using the NequIP architecture\cite{Batzner2022}, and used the resulting models to drive MD simulations in LAMMPS\cite{LAMMPS}. We used the implementations of the NequIP architecture, the training loop, and the relevant \texttt{pair style} from the \texttt{graph-pes} package to do this\cite{GraphPES}. We find that both GAP and NequIP models fitted to the same GAP-RSS dataset can qualitatively describe key structural features of liquid water (Fig.~\ref{fig:water}a--b), with the NequIP model showing some improvements for the third peak in the H--O RDF and the first peak in the O--O RDF (Fig.~\ref{fig:water}b). We emphasise that both fits could most likely be improved by using hand-crafted datasets: the aim of Fig.~\ref{fig:water} is not to benchmark specific architectures, but to test what type of practical use can be gained from an MLIP model fitted purely to automatically generated RSS data. 

The unique properties of liquid water are usually attributed to its strong hydrogen-bond network \cite{poole1994effect, kumar2007hydrogen}. Figure \ref{fig:water}c shows that both our GAP and NequIP models, fitted to the same {\tt autoplex}-generated GAP-RSS dataset, can appropriately describe the number of hydrogen bonds per molecule in liquid water in molecular-dynamics simulations, falling within the previously reported range of 3.48 to 3.84 based on different density functionals \cite{Todorova2006}. The NequIP model predicts a slightly higher average than the original GAP-RSS one, around 3.6, which is closer to the upper end of the range. We note in passing that MD simulations using random-search-based `ephemeral data-derived potentials' (EDDPs) have been presented recently \cite{Salzbrenner2023}, in that case for hydrogen diffusion in \ce{ScH12}.

Beyond liquid water, we also tested our models by predicting the energies of 52 ice crystal structures \cite{monserrat2020liquid} (structures taken from Ref.\ \citenum{monserrat2020liquid}; Fig.~\ref{fig:water}d). Here, the results from the GAP-RSS model are highly scattered, whereas the NequIP model shows improved predictive accuracy. The poor performance of the GAP model could be due to the presence of low-density phases in the dataset---which, in turn, would underscore the extrapolation capability of the NequIP model, allowing it to better handle structures that are substantially different from the training data. 

The above discussion suggests that the GAP-RSS dataset is not only effective for training a GAP model itself, but is also beneficial for use with other, more complex fitting frameworks. Future work will explore the suitability of {\tt autoplex}-generated (GAP-) RSS datasets as a pre-training task for subsequent fine-tuning of NequIP (and other MLIP) models, building on our previous study which showed that pre-training NequIP models can enhance not only numerical quality in the low-data regime, but also stability in MD simulations \cite{Gardner2024}.

\subsection*{Application to chalcogenide memory materials}

\begin{figure*}
    \centering
    \includegraphics[width=\linewidth]{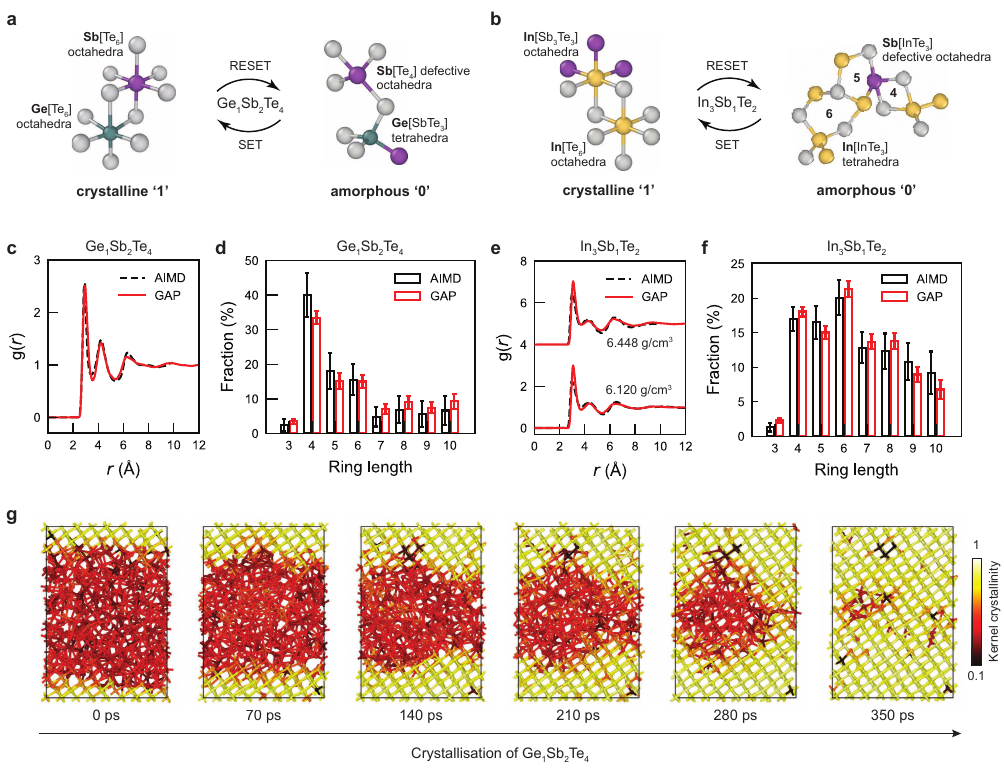}
    \caption{\textbf{Structural and dynamic properties of phase-change materials.} ({\bf a}--{\bf b}) Schematics of the structural transitions for \ce{Ge1Sb2Te4} and \ce{In3Sb1Te2}, respectively, between crystalline (`1') and amorphous (`0') phases. ({\bf c}) Radial distribution function (RDF) and ({\bf d}) distribution of shortest-path rings in \textit{a}-\ce{Ge1Sb2Te4}. ({\bf e}--{\bf f}) Same but for \textit{a}-\ce{In3Sb1Te2}. All AIMD data for \textit{a}-\ce{Ge1Sb2Te4} are sourced from the literature \cite{zhou2023device}. For \textit{a}-\ce{In3Sb1Te2}, the AIMD RDF data for a density of 6.448 $\mathrm{g\ cm^{-3}}$ are sourced from Ref.~\citenum{PhysRevB.88.174203}, while the AIMD data at 6.120 $\mathrm{g\ cm^{-3}}$ for the RDF in panel ({\bf e}) and the ring distribution in panel ({\bf f}) are obtained from our own simulations. The label `GAP' in these panels refers to our GAP-RSS-derived potentials. The bar charts in panels ({\bf d})  and ({\bf f}) display the mean values, with error bars representing the standard deviations computed across trajectories. ({\bf g}) A crystallisation simulation of \ce{Ge1Sb2Te4} using our GAP-RSS-derived potential, showing the progressive crystallisation over a simulation period of 350 ps. Atoms are colour-coded based on a kernel-based `crystallinity' measure \cite{Bartok2013, Xu2022}: yellow indicates high crystallinity, and red indicates amorphous regions.}
    \label{fig:GST}
\end{figure*}

We finally demonstrate the applicability of {\tt autoplex} to an inorganic material system of `real-world' interest. For this purpose, we focus on two ternary chalcogenides, viz.\ \ce{Ge1Sb2Te4} and \ce{In3Sb1Te2}, which are relevant for applications in phase-change memory devices \cite{Wuttig2007, Selmo2016IST, Zhang2019a}.
We have recently hand-built a GAP model for compositions along the pseudo-binary line between GeTe and \ce{Sb2Te3} (referred to as `GST' alloys)\cite{zhou2023device}, enabling an accurate description of the amorphous structure (digital `zero bits') and its crystallisation (`0 $\rightarrow$ 1'). One of the challenges in modelling amorphous GST is the formation of tetrahedral structural motifs (Fig. \ref{fig:GST}a), which are relevant for ageing phenomena: the amount of those tetrahedra has been argued to change over time, affecting the stability of the zero bits \cite{Raty2015}. However, this hand-crafted GAP model (denoted `GST-GAP-22') took months to complete, involving multiple runs of domain-specific iterations to cover the structural complexity of the Ge--Sb--Te system \cite{zhou2023device}. This makes \ce{Ge1Sb2Te4} an excellent candidate for testing our automated, RSS-based workflows in computational practice.

Compared to GST, \ce{In3Sb1Te2} (`IST' in the following) is a rather unconventional phase-change material, with structural building blocks slightly different from those of GST. The latter alloys structurally resemble their constituent binary phases (viz.\ GeTe and \ce{Sb2Te3}), and they can take disordered and defective rocksalt-like metastable structures (see Ref.~\citenum{Deringer2015a} and references therein). The relevant ternary In--Sb--Te compound, viz.\ \ce{In3Sb1Te2}, crystallises in a disordered rocksalt-type structure with no substantial amount of cation vacancies \cite{Schroeder2013}. Unlike GST, which does contain such vacancies, IST has all cation sites fully occupied by \ce{In} atoms (Fig. \ref{fig:GST}b), with no notable cation disorder. Instead, its structural complexity arises from {\em anion} disorder, as \ce{Sb} and \ce{Te} share the same substructure. A previous AIMD study revealed that amorphous IST exhibits a large number of four-, five-, and six-membered rings \cite{PhysRevB.88.174203} (Fig. \ref{fig:GST}b, right), indicating that its medium-range order is more complex than that of amorphous GST (where four-membered rings are predominant).

In the context of MLIP development, IST is an example of a less widely explored material: there are not as many AIMD studies as for GST, nor is there an existing potential model to our knowledge. We argue that instead of manually constructing a training dataset, the user can now use automated approaches, such as the one in {\tt autoplex}, to study less-common functional materials, at least as a starting point.

Figure~\ref{fig:GST}c--g characterises the performance of GAP-RSS models for both phase-change materials. We first quantified the local structural properties by computing RDFs (Fig.~\ref{fig:GST}c and Fig.~\ref{fig:GST}e), and then calculated ring statistics which allow us to assess the medium-range order in those structures (Fig.~\ref{fig:GST}d and Fig.~\ref{fig:GST}f). The performance of these models is encouraging, given how inexpensive it has been to fit compared to existing MLIPs for Ge--Sb--Te compounds (e.g., the one of Ref.~\citenum{zhou2023device}, fitted to a large set of DFT data including domain-specific configurations). Moreover, it shows that our approach not only works well for materials with established domain knowledge but also performs effectively for materials with limited domain knowledge.

We finally ran a GAP-driven MD simulation of the crystallisation process in \ce{Ge1Sb2Te4} (Fig.~\ref{fig:GST}g), corresponding to the SET operation in memory devices (`0 $\rightarrow$ 1'). We used a kernel similarity metric \cite{Bartok2013} to quantify the gradual structural ordering process, as in our previous work (Ref.~\citenum{Xu2022}). The simulation shows a rapid growth proceeding at the amorphous--crystalline interface, leading to the formation of a largely ordered, defective rocksalt-like structure. We note that this crystallisation simulation using GAP was essentially completed after 350 ps, more quickly than what was seen in {\em ab initio} (DFT-based) MD simulations \cite{Xu2022} and using the hand-crafted GST-GAP-22 potential \cite{zhou2023device}. However, the former approach is computationally highly demanding\cite{Xu2022}, and the latter is a specialised MLIP that has been deliberately trained on `domain-specific' configurations that correspond to the intermediate steps between fully amorphous and fully recrystallised Ge--Sb--Te materials. Our GAP-RSS based model, used to drive the simulations characterised in Fig.~\ref{fig:GST}g, is trained in a much simpler way, and using only relatively small-scale configurations---perhaps it could serve as a starting point for constructing subsequent, more complex MLIP training datasets. Further tests for other phase-change memory materials are planned for the future.

\section*{Discussion}

Automation is one of the major open challenges in ML potentials for materials and poised to accelerate their development into general mainstream simulation models.
We have here shown how iterative exploration and MLIP fitting, initially proposed within the GAP-RSS framework \cite{Deringer2018}, can be automated substantially and at scale by integration with existing software ecosystems.
The resulting {\tt autoplex} code is openly available and free to use, and we expect that in the long run, it will develop into a computational ecosystem of its own for the next generation of ML-driven materials modelling.

Methodologically, our work contributes to addressing the wider-ranging open question about how MLIPs are best developed and used going forward. Random searching provides a core approach to generating robust potentials, and our present results suggest a remarkable amount of stability that can be gained from RSS using small cells alone: therefore ML-driven iterative RSS appears to emerge as a standard technique for at least providing a starting point for potential fitting \cite{Deringer2020c}.
We have previously pointed out that RSS datasets can constitute useful benchmark tasks for evaluating MLIP models \cite{Morrow2023, ThomasduToit2024}, and we expect growing usefulness of this aspect as systematic benchmarking becomes more important in the community.

In the years ahead, we expect that automated approaches to dataset construction---including the one in {\tt autoplex} we have presented here---will play an increasingly important role in the field. With these efforts, and together with the demonstrated capabilities of advanced MLIP fitting architectures \cite{Merchant2023, zhang2024dpa2largeatomicmodel, Batatia2023, Yang2024}, universal ML models for atomistic simulations could become widely established---which could make ML-driven modelling the genuine default in the field, just like direct quantum-mechanical modelling has been the default for many years.

\clearpage
\setstretch{1.1}

\section*{Methods}

\subsection*{ML potentials}

MLIPs represent a given quantum-mechanical potential-energy surface. The models used herein are based on a local (atom-wise) decomposition of the total energy \cite{Behler2007, Bartok2010},
\begin{equation}
    \hat{E} = \sum_{i} \hat{\varepsilon}({\vec x}_{i}),
\end{equation}
where the atomic energies are learned as a function of the atom's local environment, expressed through a general descriptor, $\vec{x}_{i}$, and the ML-predicted total energy, $\hat{E}$, is obtained by summing over the per-atom contributions.

For most of the present study, we used the Gaussian Approximation Potential (GAP) framework \cite{Bartok2010}, together with the SOAP descriptor \cite{Bartok2013} to featurise atomic environments.
However, the approaches are more general, and we include interfaces to the {\tt ACEpotentials.jl} \cite{Witt2023}, NequIP \cite{Batzner2022}, M3GNet \cite{Chen2022}, and MACE\cite{MACE2022} codes in the current public version of {\tt autoplex} as well.
We refer to the original literature for details of the methods.

\subsection*{RSS and iterative fitting}

Random searching is an established approach to exploring configurational spaces and has seen large success in the context of the AIRSS framework by Pickard and Needs \cite{Pickard2006, Pickard2011}.
In Ref.~\citenum{Deringer2018}, it was proposed to combine RSS exploration with the fitting of potential-energy surfaces, gradually improving the models through iterative training (which had already been in itself a standard approach using iterative MD \cite{Sosso2012}) and not requiring any DFT-based relaxation but only single-point DFT evaluations, running all relaxations with GAP models instead.
Note that the combination of structural searches and iterative MLIP fitting is not at all restricted to GAP or (AI)RSS---it has been demonstrated for other frameworks as well \cite{Tong2018, Podryabinkin2019}.

The approach was extended in Ref.~\citenum{Bernstein2019b} by including appropriate selection steps (both structurally and energetically based). Subsequent work introduced the {\tt wfl} software \cite{gelvzinyte2023wfl}. The latter work is different from ours in that it runs workflows through a custom implementation, whereas {\tt autoplex} interfaces to {\tt atomate2} and its diverse set of DFT-based workflows with default inputs and other software where possible.

Within {\tt autoplex}, we have incorporated several methodological features. Our framework now supports multiple sets of {\tt buildcell} input parameters, which define the scope of the RSS search and directly impact the diversity of the generated structures. Furthermore, we have implemented a Hookean repulsion force to prevent atoms from approaching each other too closely, which can help to produce more physically plausible configurations.

\subsection*{Automation}

In the present work, we focus on automating RSS and RSS-driven iterative potential fitting. We describe a software implementation that is connected to the {\tt atomate2} ecosystem. {\tt atomate2} is a library of computational materials science workflows that have mostly been developed in the context of the Materials Project \cite{Jain2013}. 
As in {\tt atomate2}, we rely on {\tt jobflow}\cite{rosen_jobflow_2024} for writing workflows and {\tt jobflow-remote} or {\tt fireworks}\cite{jain_fireworks:_2015} for executing workflows and scheduling computing tasks.
For handling DFT inputs and outputs and interfaces with ML potentials, we rely on {\tt pymatgen} \cite{Ong2013} and the Atomic Simulation Environment (ASE) \cite{HjorthLarsen2017}.

For automatic DFT computations, we use existing {\tt atomate2} workflows but set default parameters suitable for high-accuracy MLIPs. We furthermore automated in-between between data generation on the one hand (DFT, typically), and fitting MLIPs on the other hand. We define both as methods-agnostic: for the DFT part, the automation is handled by the {\tt atomate2} etc.\ frameworks, for the fitting part, we provide direct interfaces to relevant software.

\subsection*{Technical details}

Reference data were obtained using projector augmented-wave (PAW) potentials \cite{Blochl1994, Kresse1999} as implemented in the Vienna {\em Ab Initio} Simulation Package (VASP) \cite{Kresse1996, Kresse1999}. 
To describe the effects of exchange and correlation, we used the Perdew--Burke--Ernzerhof (PBE) functional \cite{Perdew1996}, its revised parameterisation for solids (PBEsol) \cite{Perdew2008}, as well as the Strongly Constrained and Appropriately Normed (SCAN) functional \cite{Sun2015}, depending on the specific material system to be studied.
The DFT settings were similar to previous work for silicon \cite{Bernstein2019b}, Ti--O \cite{lyle2015prediction}, \ce{SiO2} (Ref.~\citenum{Erhard2022}), water \cite{PhysRevLett.126.236001}, and phase-change materials \cite{zhou2023device} (here using the same settings for GST and IST), and adapted where appropriate.
Structures were visualised using OVITO \cite{Stukowski2009}.

\section*{Data availability}

Data supporting this work, including raw data and Python notebooks to reproduce the plots, will be made available via GitHub upon journal publication.

\section*{Code availability}

The {\tt autoplex} software is openly available at \href{https://github.com/autoatml/autoplex}{\tt https://github.com/autoatml/autoplex}.
The code is under ongoing development; a copy of the version used for the results presented herein ({\tt v0.0.7}) is deposited in Zenodo (\href{https://doi.org/10.5281/zenodo.14169361}{\tt https://doi.org/10.5281/zenodo.14169361}).

\clearpage

\section*{Acknowledgements}
We thank L.-B. Pasca and L. Wu for help with early tests.
This work was supported through a UK Research and Innovation Frontier Research grant [grant number EP/X016188/1].
J.D.M. acknowledges funding from the EPSRC Centre for Doctoral Training in Inorganic Chemistry for Future Manufacturing (OxICFM), EP/S023828/1.
We are grateful for computational support from the UK national high performance computing service, ARCHER2, for which access was obtained via the UKCP consortium and funded by EPSRC grant ref EP/X035891/1. 
Additionally, we acknowledge the Gauss Centre for Supercomputing e.V. (www.gauss-centre.eu) for funding this project by providing generous computing time on the GCS Supercomputer SuperMUC-NG at Leibniz Supercomputing Centre (www.lrz.de) (project pn73da) that enabled testing of the implementations in {\tt autoplex}.

\section*{Author contributions}
Y.L. and J.D.M. developed the RSS automation code in {\tt autoplex}. Y.L., C.E., N.L.F., A.A.N., and J.G. are the core {\tt autoplex} code developers and maintainers at the time of this writing. Y.L., J.L.A.G., N.L.F., and Y.Z. carried out numerical experiments. J.G. and V.L.D. designed and supervised the research. Y.L. and V.L.D. drafted the manuscript, and all authors contributed to the final version.

\end{document}